\documentclass[3p]{elsarticle}
\usepackage{amsfonts}
\usepackage{mathrsfs}
\usepackage{amsmath}
\usepackage{xcolor}
\usepackage{amssymb}
\usepackage{bm}
\usepackage{feynmp}
\usepackage{extarrows}
\usepackage{slashed}
\usepackage{graphicx}
\usepackage{subfigure}

\usepackage[linktocpage=true,bookmarks=true,bookmarksnumbered=true,breaklinks=true,pdfpagemode=Fullscreen,pdfstartview=FitBH]{hyperref}

\newcommand{\FR}[2]{\displaystyle\frac{\,{#1}\,}{#2}}
\newcommand{\fr}[2]{\mbox{$\frac{\,{#1}\,}{#2}$}}
\newcommand{\n}{\nonumber}
\renewcommand{\rm}{\mathrm}
\def\bge{\begin{equation}}
\def\ede{\end{equation}}
\def\bga{\begin{aligned}}
\def\eda{\end{aligned}}
\newcommand{\beq}{\begin{equation}}
\newcommand{\eeq}{\end{equation}}
\newcommand{\bq}{\begin{equation}}
\newcommand{\eq}{\end{equation}}
\newcommand{\ba}{\begin{array}}
\newcommand{\ea}{\end{array}}
\newcommand{\beqa}{\begin{eqnarray}}
\newcommand{\eeqa}{\end{eqnarray}}
\newcommand{\beqs}{\begin{subequations}}
\newcommand{\eeqs}{\end{subequations}}

\def\dis{\displaystyle}

\def\({\left(}
\def\){\right)}

\def\End{\end{document}}
\newcommand{\order}[1]{\mathcal{O}({#1})}

\def\RE{\Re\mathfrak{e}}
\def\IM{\Im\mathfrak{m}}

\def\leqq{\leqslant}

\def\dif{\partial}

\def\ka{\kappa}

\def\ga{\gamma}
\def\de{\delta}
\def\ep{\epsilon}
\def\lam{\lambda}
\def\rh{\rho}
\def\si{\sigma}

\def\di{\mathrm{d}}
\def\D{\mathrm{D}}
\def\T{\mathrm{T}}

\def\pd{\partial}
\def\ld{{\mathscr{L}}}

\def\ra{\rangle}

\def\tr{\mathrm{\,tr\,}}

\def\to{\rightarrow}

\def\ii{\mathrm{i}}

\def\ECM{E_{\text{cm}}}
\def\T{{\mathcal{T}}}
\def\C{{\mathrm{C}}}
\def\el{{\mathrm{el}}}
\def\inel{{\mathrm{inel}}}
\def\cut{\Lambda_{\textrm{UV}}^{}}

\begin{document}

\title{{\bf Spontaneous Spacetime Reduction and \\
            Unitary Weak Boson Scattering at the LHC}}

\author[]{{\sc Hong-Jian He}$\,^{a,b,c}$ ~and~
          {\sc Zhong-Zhi Xianyu}$\,^a$}


\address[tuhep]{Institute of Modern Physics and Center for High Energy Physics,
                Tsinghua University, Beijing 100084, China}
\address[puhep]{Center for High Energy Physics, Peking University, Beijing 100871, China}
\address[kavli]{Kavli Institute for Theoretical Physics China, CAS, Beijing 100190, China}

\begin{abstract}
Theories of quantum gravity predict spacetime dimensions to become reduced at high energies,
a striking phenomenon known as spontaneous dimensional reduction (SDR).  We construct an
effective electroweak theory based on the standard model (SM) and incorporate the TeV-scale 
SDR, which exhibits good high energy behavior and ensures the unitarity of weak gauge boson
scattering. This also provides a natural solution to the hierarchy problem in the presence 
of scalar Higgs boson. We demonstrate that this model predicts unitary longitudinal 
weak boson scattering, and can be discriminated from the conventional 4d SM by the $WW$ 
scattering experiments at the CERN LHC.
\\[2mm]
PACS numbers: 04.60.-m, 11.80.-m, 12.60.-i
\hfill
Phys.\ Lett.\ B (in Press)  
\end{abstract}

\maketitle


\vspace{4mm}
 \section{Introduction}

 Spontaneous dimensional reduction (SDR) \cite{SDR} is a striking phenomenon, showing that
 the spacetime dimensions effectively equal $3\!+\!1$ at low energies, but get reduced
 toward $1\!+\!1$ at high energies. This is predicted
 by a number of quantum gravity approaches \cite{SDR}, including
 the causal dynamical triangulation, the exact renormalization group method,
 the loop quantum gravity, the high-temperature string theory,
 and Ho\v{r}ava-Lifshitz gravity, etc.
 The SDR is expected to greatly improve the ultraviolet (UV) behavior of the
 standard model (SM) of particle physics.
 Recently, this property of the quantum gravity
 has called phenomenological interests, and has found applications
 to astrophysics and collider phenomenologies in a different context\,\cite{TeV-app}.
 In addition, some hints of a TeV scale SDR have been noted\,\cite{TeV-app} from the
 observations of alignment of high energy cosmic rays\,\cite{SDR-cosmicray}.

\vspace*{1mm}

 The $SU(2)_L^{}\otimes U(1)_Y^{}$ gauge structure of the SM is
 well established for electroweak forces.
 In the conventional SM, it is linearly realized and spontaneously broken
 by the Higgs mechanism\,\cite{HM}, leading to massive weak gauge bosons $(W^\pm,Z^0)$.
 A Higgs boson is predicted and is crucial for the renormalizability
 and unitarity of the SM.  Recently, ATLAS and CMS collaborations
 have found signals for a new particle with mass around 125\,GeV at the LHC,
 which are somewhat different from the SM Higgs boson (especially in the diphoton mode),
 although still consistent with the SM expectations
 within about $2\sigma$ statistics \cite{LHCnew}.
 Hence, the true mechanism of electroweak symmetry breaking is awaiting
 further explorations at the LHC, and the possible new physics beyond the SM Higgs boson
 is highly anticipated. Given the fact that no other new particles have been detected 
 so far, it is tantalizing to explore alternative new physics sources at the TeV scale,
 beyond the conventional proposals such as extra dimensions, supersymmetry and 
 strong dynamics at the TeV scale.

\vspace*{1mm}

 In this Letter, we will explore the quantum gravity effect of SDR at TeV scale,
 and study its applications to the electroweak sector of the SM.
 We conjecture that the TeV Scale SDR can play a key role to unitarize
 weak gauge boson scattering in the theory without or with a light Higgs boson.
 As a first example, we will show that the perturbative unitarity is maintained by
 the TeV scale SDR in scenarios without a Higgs boson, where the recently observed
 125\,GeV boson \cite{LHCnew} can be something else, such as a dilaton-like
 particle \cite{dilaton}.
 Without a Higgs boson, the SM electroweak gauge symmetry
 $SU(2)_L^{}\otimes U(1)_Y^{}$  becomes nonlinearly realized\,\cite{App}
 and the three Goldstone bosons are converted to the longitudinal polarizations of
 $(W^\pm ,\,Z^0)$ after spontaneous symmetry breaking.
 Such a minimal Higgsless SM loses traditional renormalizability \cite{App} and
 violates unitarity of weak boson scattering at the TeV scale \cite{TreeU},
 hence it is incomplete. We show that the TeV-scale SDR can provide a
 new way to unitarize the $WW$ scattering, and will be discriminated from the SM
 at the LHC.

\vspace*{1mm}

 Then, we study the SM with a non-standard Higgs boson of mass around $125$\,GeV
 under the TeV-scale SDR (called the Higgsful SM).
 We will show that the corresponding $WW$ scattering cross
 sections become unitary at TeV scale under the SDR,
 but exhibit different behaviors from the conventional 4d SM.
 We note that different ways of unitarizing the longitudinal $WW$ scattering
 around TeV scale reflect the underlying mechanisms
 of electroweak symmetry breaking (EWSB),
 and will be discriminated by the $WW$ scattering experiments
 as a key task of the LHC \cite{WW-rev}.


\vspace*{2mm}
\section{The TeV Scale SDR}

 Despite lacking a full theory of quantum gravity that could precisely describe the SDR,
 we are modest and approach this problem by using the
 {\it effective theory formulation} \cite{EFT}.
 In particular, to mimic the result from the causal dynamical triangulation \cite{ambjorn},
 we parameterize the spacetime dimension $\,n=n(\mu)$\, as a smooth function of
 the energy scale $\mu$ (which we call the dimensional flow by following Calcagni\,\cite{FST}),
 such that $\,n(\mu)\to 4$\, under $\,\mu\to 0\,$
 in the infrared region as supported by all low energy experiments,
 and $\,n(\mu )\to 2$\, at a certain UV scale $\,\cut$\,.\,
 We can make a simple choice for the dimensional flow,
 \bge
   \label{eq:DFansatz}
    n(\mu) ~=~ 4-2\(\FR{\mu}{\cut}\)^{\!\gamma},  ~~~~~~~  (\,\mu\leqq\cut\,)\,,
 \ede
 where the index $\,\gamma > 1\,$ is a model-dependent parameter, determined
 by the nonperturbative dynamics of quantum gravity.
 As simple realizations, we may set, $\,\gamma=2\,$ or 1.5\,.\,
 Before finding a unique full theory of the quantum gravity,
 other variations of (\ref{eq:DFansatz}) are possible\,\cite{ambjorn,FST},
 but this will not affect the main physics features of the present analysis.
 An easy choice for $\cut$ would be the Planck scale.
 But it is a very interesting and intriguing possibility
 that the nonperturbative dynamics of quantum gravity
 drives $\cut$ down to $\order{\text{TeV}}$ \cite{TeV-app}.
 If this happens, a number of difficulties associated with the EWSB and
 $W/Z$ mass-generations in the SM can be resolved without introducing additional
 {\it ad hoc} hypothetical dynamics.

\vspace*{1mm}

  When the quantum gravity effects show up at the TeV scale, they will induce effective
  operators causing sizable anomalous Higgs couplings to $WW\;(ZZ)$ and fermions
  in the low energy effective theory. This will violate perturbative unitarity
  at TeV energy scale in the conventional 4d setup \cite{HVV-0}\cite{He:2002qi}.
  However, we show that under the TeV-scale SDR, the weak boson scattering amplitudes
  will still be unitarized through the reduction of spacetime dimensions.
  Furthermore, the presence of TeV scale SDR also provides a natural solution
  to the hierarchy problem since the 4d quadratically divergent radiative corrections
  to Higgs boson mass is rendered to be logarithmic in $\,n=2$\, spacetime and thus harmless.


\vspace*{2mm}
\section{The Standard Model with SDR}

 As an effective theory description of the SDR,
 we encode the information of dimensional flow $\,n=n(\mu)$\,
 into the measure of spacetime integral $\,\di\rho$\,,\,
 and replace all integral measure $\,\di^4x\,$
 in the action functional by $\,\di\rho$\,.\,
 A rigorous mathematical construction of $\,\di\rh\,$
 is given by Ref.\,\cite{FST},
 but the detail is not needed here.
 All we need to know is that the mass-dimension of this measure is
 $\,[\di\rh]=-n$\,,\, where $\,n=n(\mu)$\,
 is the dimensional flow in Eq.\,(\ref{eq:DFansatz}).
 It is enough to define the measure $\,\di\rh\,$
 formally by $\,\di^nx$\,,\, with $\,n\,$ a scale-dependent quantity.
 Thus, we can write down the action of the theory,
 \,$S=\int\!\di^nx\,\ld = \int\!\di^nx\,(\ld_G+\ld_F)$\,,\,
 where $\,\ld_G\,$ and $\,\ld_F\,$ are the gauge and fermion parts
 of the SM Lagrangian.
 We will focus on the gauge sector for the current study.
 We first consider the gauge Lagrangian with Higgs boson removed,
 \beqa
 \label{eq:L}
   \ld_G^{} &\!=\!&  \dis
   -\fr{1}{4}W_{\mu\nu}^aW^{\mu\nu a}
   -\fr{1}{4}B_{\mu\nu}^{}B^{\mu\nu}_{}
   \dis +M_W^2 W_\mu^+W^{-\mu}+ \fr{1}{\,2\cos^2\theta_w\,}M_W^2Z_\mu^{} Z^\mu_{} \,,
   ~~~~
 \eeqa
 where the gauge field strength  
 $\,W_{\mu\nu}^a = \pd_\mu^{} W_\nu^a-\pd_\nu^{} W_\mu^a + g\ep^{abc}W_\mu^b W_\nu^c$\, 
 and $\,B_{\mu\nu}^{} =  \pd_\mu^{} B_\nu^{}-\pd_\nu^{} B_\mu^{}\,$.\,
 In the above, $\,\theta_w=\arctan (g'/g)\,$ 
 represents the weak mixing angle and connects the gauge-eigenbasis
 \,$(W^3_\mu,\,B_\mu^{})$\, to the mass-eigenbasis \,$(Z^0_\mu,\,A_\mu^{})$\,.\,
 Eq.\,(\ref{eq:L}) contains $(W,Z)$ mass terms in unitary gauge 
 and can be made gauge-invariant in the nonlinear realization 
 of $SU(2)_L^{}\otimes U(1)_Y^{}$ gauge symmetry,
 \beqa
 \label{eq:L-Sigma}
 \ld_\Sigma^{} ~=~ \fr{1}{4}v^2\tr\!\left[(\D^\mu\Sigma)^\dag(\D_\mu\Sigma)\right] ,
 \eeqa
 where $\,\D_\mu\Sigma = \pd_\mu^{}\Sigma
        + \fr{\ii}{2} g{W}_\mu^a\tau^a\Sigma-\fr{\ii}{2}g'B_\mu\Sigma\,\tau^3 \,$,\,
 and  $\,\Sigma = \exp [\ii\tau^a\pi^a/v]$\, with $\{\pi^a\}$ the Goldstone bosons.
 Eq.\,(\ref{eq:L-Sigma}) gives, $\,M_W=\fr{1}{2}gv\,$,\,
 where the parameter $\,v\,$ will be fixed by the low energy Fermi constant 
 $\,G_F^{}=(\sqrt{2}v^2)^{-1}_{}$\,.\,
 The Lagrangian (\ref{eq:L}) derives directly from the SM in unitary gauge
 after removing the Higgs boson; while Eq.\,(\ref{eq:L-Sigma}) is just the lowest order
 electroweak chiral Lagrangian of the SM \cite{App}.

\vspace*{1mm}

 We can further embed the Higgs boson as a singlet scalar $\,h^0$\, in this formulation
 by extending the Lagrangian (\ref{eq:L-Sigma}) as follows \cite{HVV-0}\cite{He:2002qi},
 \beqa
 \label{eq:L-h-Sigma}
 \ld_H^{} &\!=\!&
 \frac{1}{4}\(v^2+2\ka v h +\ka' h^2\)
 \tr\!\!\left[(\D^\mu\Sigma)^\dag(\D_\mu\Sigma)\right]
 \n\\[1mm]
 &&
 +\frac{1}{2}\dif_\mu^{}h\dif^\mu h
 -\frac{1}{2}M_h^2h^2 - \frac{\lambda_3^{}}{3!}vh^3
 +\frac{\lambda_4^{}}{4!}h^4 \,,~~~~~
 \eeqa
 where $\,\Delta\ka \equiv\kappa-1\,$ and $\,\Delta\ka' \equiv\kappa'-1\,$ denote
 the anomalous gauge couplings of the Higgs boson $\,h^0$\, with $\,WW\,$ and $\,ZZ\,$.\,
 The conventional 4d SM is just a special case, $\,\Delta\ka=\Delta\ka'=0$\,,\,
 and $\,\lambda_{3}^{}=\lambda_{4}^{}=\lambda_0^{}=3M_h^2/v^2\,$,\,
 in the general effective Lagrangian (\ref{eq:L-h-Sigma}).
 For non-zero anomalous couplings
 $\,\Delta\ka,\Delta\ka'\neq 0$\, and/or
 $\,\lambda_{3}^{},\lambda_{4}^{} \neq \lambda_0^{}\,$,\,
 as induced by the effects of TeV scale SDR,\,
 the scalar field $h^0$ becomes a non-standard Higgs boson.
 We will study how to discriminate such as non-SM Higgs particle from the
 conventional 4d SM in Sec.\,5.

\vspace*{1mm}

 We also note that our Lagrangian $\ld$ is {manifestly Lorentz invariant} and thus
 all particles' dispersion relations remain unchanged, as in \cite{FST}.
 This is because our formulation is based on the framework of \cite{FST},
 where it is shown that the action can be constructed in such a way that
 the Lagrangian density $\ld$ lies in $3\!+\!1$ dimensional spacetime and
 respects the $(3\!+\!1)$d Poincar\'e symmetry, while the effect of SDR 
 is fully governed by a properly defined integral measure $\di\rho$\,.\,
 In such a scenario, scalar, spinor and vector fields are linear representations of
 (3+1)d Lorentz group $SO(3,1)$ (up to a gauge transformation for gauge fields).
 Practically, this is similar to the conventional dimensional reduction regularization
 method \cite{DRED}, which maintains the 4d Lorentz symmetry and continues physics to
 $\,n < 4\,$.
 Thus, our model is free from Lorentz-violation constraints
 in the cosmic ray observations and collider experiments at the tree level.
 In addition, our present study focuses on the scattering of longitudinal weak bosons,
 and their amplitudes are equivalent to that of the corresponding Goldstone bosons
 at high energies according to the equivalence theorem \cite{ETrev}.
 The amplitudes of scalar
 Goldstone bosons are cleanly defined in general dimension $\,n\,$.\,
 Manipulations of scalar fields do not involve contracting or counting Lorentz indices,
 and thus do not rely on details of realizing the SDR.

\vspace*{1mm}

 In general $n$-dimensional spacetime, the action functional should remain dimensionless.
 Hence the Lagrangian has the mass-dimension $\,[\ld]=n$\,,\,
 and the gauge coupling $\,g\,$  has mass-dimension $\,[g]=(4-n)/2$\,.\,
 We can always define a new dimensionless coupling $\,\tilde{g}\,$
 and transfer the mass-dimension of $\,g\,$ to another mass-parameter.
 Since the gauge coupling $\,g$\, in (\ref{eq:L}) becomes super-renormalizable
 for $\,n<4\,$ and thus insensitive to the UV, it is natural to scale $\,g\,$
 by the $W$ mass $M_W$ [which is the only dimensionful parameter of the
 Lagrangian (\ref{eq:L}) in 4d],
 \beqa
 \label{eq:gt}
 g ~=~ \tilde{g}\, M_W^{(4-n)/2} \,,\,
 \eeqa
 with $\,\tilde{g}\,$ being dimensionless.
 The value of  $\,\tilde g$\, is given by that of $\,g\,$ at $\,n=4$\,.\,
 We will concentrate on the tree-level analysis in this study,
 so the coupling $\,\tilde{g}\,$ is a scale-independent constant.

\vspace*{1mm}

 In fact, the scaling (\ref{eq:gt}) is well justified for more reasons.
 We may easily wonder why we could not use the UV cutoff $\cut$ in the scaling of
 $\,g\,$ as a replacement of the infrared mass-parameter $M_W$ of the theory.
 This is because in spacetime dimension $\,n<4$\,,\,
 the gauge coupling $g$ is super-renormalizable with positive mass-dimension
 $\,[g]=(4-n)/2 > 0\,$.\,
 Such a super-renormalizable coupling must be insensitive
 to the UV cutoff of the theory, contrary to a non-renormalizable coupling
 with negative mass-dimension and thus
 naturally suppressed by negative powers of the UV cutoff $\cut$
 ({\it e.g.,} in $n>4$\, or in association with certain higher-dimensional operators).
 It is easy to imagine that for a super-renormalizable theory
 in dimension $\,n<4\,$,\, if its coupling $\,g\,$ were scaled as
 $\,g=\tilde{g}\Lambda_{\rm UV}^{(4-n)/2}$,\,
 it would even make tree-level amplitude UV divergent and blow up as $\cut\to\infty$;
 this is clearly not true. On the other hand, it is well-known that a non-renormalizable
 coupling $\,g\,$ with negative mass-dimension $\,[g]\equiv -p <0$\,
 should be scaled as $\,g=\tilde{g}/\Lambda_{\rm UV}^{p}\,$,\,
 and thus the tree-level amplitude
 naturally approaches zero when $\,\cut\to\infty$,\, as expected.

\vspace*{1mm}

 Since spacetime dimension flows to $\,n=2\,$ in the UV limit, we observe that
 the theory (\ref{eq:L}) is well-behaved at high energies.
 This is because all gauge couplings of the Lagrangian (\ref{eq:L})
 in $\,n<4\,$ dimensions becomes super-renormalizable, and the gauge boson propagators scale
 as $1/p^{2}$ in high momentum limit under the $R_\xi^{}$ gauge-fixing.
 So we only concerns about gauge boson mass-terms in (\ref{eq:L}) or (\ref{eq:L-Sigma}),
 which is the origin of nonrenormalizability and unitarity violation in 4d.
 But, in our construction the spacetime dimension flows to $\,n=2\,$
 in high energy limit where (\ref{eq:L-Sigma}) just describes a
 2d gauged nonlinear sigma model and is renormalizable, as is well known.

\vspace*{1mm}

 We also note that in $\,n<4\,$ dimensions gauge bosons can acquire masses via new mechanisms
 other than the Higgs mechanism.
 For instance, in the 2d Schwinger model \cite{schwinger},
 radiative corrections to the vacuum polarization from
 a massless-fermion loop generate a nonzero photon mass,
 $\,m_\ga^{} = \frac{e}{\sqrt{\pi}\,}$\,.\,
 Also, the 3d Chern-Simons term induces a topological mass for the
 corresponding gauge field \cite{deser}, $\,m_{\text{cs}}^{} = \ka \,e^2\,$.\,
 Hence, it is natural to have an explicit
 mass-term of vector boson in a lower dimensional field theory.
 We will further demonstrate below that such a mass-term is indeed harmless in
 a Higgsless SM with the TeV scale SDR, and the unitarity of
 high energy longitudinal $WW$ scattering is ensured.


\vspace*{2mm}
\section{Longitudinal Weak Boson Scattering under SDR}

 As a simple illustration of the unitarization mechanism of longitudinal $WW$
 scattering under the SDR, we first present the analysis in
 the SM without Higgs boson (called the Higgsless SM and denoted by HLSM-SDR).
 It is noted that such a scenario can be consistent with the current LHC data since the
 125\,GeV new boson may be something else, such as a dilaton-like particle\,\cite{dilaton}.
 The effect of SDR is most clearly seen in this case. After this,
 we will further extend this mechanism to the Higgsful SM under the SDR
 (including a Higgs boson and called the HFSM-SDR), in the next section.

\vspace*{1mm}

  The minimal 4d Higgsless SM violates unitarity at TeV scale,
  because the SM Higgs boson plays the key role to unitarize the bad high energy behaviors
  of the longitudinal $WW$ scattering.
  For instance, without Higgs boson, the amplitude of
  $\,W_L^+W_L^-\to Z_L^0Z_L^0$\, has non-canceled $E^2$ term,
  \beqa
  \mathcal{T}_{\text{HL}} ~=~ {g^2\ECM^2}/(4M_W^2)+\order{\ECM^0}\,,\,
  \eeqa
  where $\ECM$ is the c.m.\ energy.
  This bad $E^2$ behavior leads to unitarity violation at TeV scale.
  In contrast, for the conventional 4d SM, this $E^2$ term is exactly canceled by
  the contribution of the $s$-channel Higgs-exchange, which is
  the key to ensure the SM unitarity.

\vspace*{1mm}

 In lower dimensions, the longitudinal amplitudes remain the same as in 4d.
 But, we observe that the form of partial wave expansion changes,
 due to the phase-space reduction for final state.
 Hence, the $E^2$-cancellation described above is no longer essential for
 ensuring the unitarity.
 This is an essential feature of the unitarization mechanism through SDR, i.e.,
 the $WW$ scattering amplitudes remain unitary at high energies under SDR,
 even without a Higgs boson.

\vspace*{1mm}

 To be explicit, we recall that unitarity condition for $S$-matrix
 arises from probability conservation,  $\,SS^\dag=S^\dag S=1$\,.\, This leads to $\,\mathscr{T}^\dag\mathscr{T}=2\IM\mathscr{T}$,\, where $\,\mathscr{T}\,$
 is defined via $\,S=1+\ii \mathscr{T}$,\, and is related to the amplitude $\T$ via $\,\mathscr{T}=(2\pi)^n\de^n(p_f^{}-p_i^{})\T$\, with $p_i^{}$ ($p_f^{}$)
 the total momentum of the initial (final) state.
 For $\,2\to 2$\, scattering, $\T$ depends only on the c.m.\ energy $\ECM$
 and scattering angle $\,\theta$\,.\,
 Thus, in this case we can always expand $\,\T(\ECM,\theta)$\,
 in terms of partial waves $\,a_\ell^{}(\ECM)$\, for $\,n >3\,$ dimensions,
 \vspace*{-2mm}
 \begin{align}
   \label{eq:PW}
   & \T ~=~ \lam_n \ECM^{4-n} \sum_{\ell}\!
   \FR{1}{N_\ell^\nu}\C_\ell^\nu(1)
   \C_\ell^\nu(\cos\theta)\,a^{}_\ell \,,
   \n\\[-2.5mm]
   \\[-2.5mm]
   & a_\ell^{} ~=~ \FR{\ECM^{n-4}}{\lam_n\C_\ell^\nu(1)}
   \!\int_0^\pi\!\!\!\!\di\theta\sin^{n-3}\theta \C_\ell^{\nu}(\cos\theta)\,\T \,,
  \n
  \\[-7mm]
  \n
 \end{align}
 with $\,\lam_n^{} = 2(16\pi)^{n/2-1}_{}\Gamma(\fr{n}{2}-1)$,~ $\nu=\fr{1}{2}(n-3)$,~ $N_\ell^\nu=\FR{\pi\Gamma(\ell\!+\!2\nu)}{2^{2\nu-1}\ell!(\ell\!+\!\nu)\Gamma^2(\nu)}$,\,
 and $\,\C_\ell^\nu(x)\,$ is the Gegenbauer polynomial
 of order $\,\nu\,$ and degree $\,\ell\,$.\,
 This partial wave expansion holds for $\,n>3\,$
 because the eigenfunctions of rotation generators (namely the Gegenbauer function)
 are not well defined below $\,n=3\,$.\,
 The appearance of factor $\,\ECM^{4-n}$\, in the expansion of $\,\T\,$
 is expected, since the $S$-matrix of $\,2\to 2\,$ scattering has a mass-dimension
 $\,4-n\,$ in $n$ dimensions, and the partial wave amplitude $a_\ell^{}$ is
 dimensionless by definition.
 Then, we can derive unitarity conditions for the elastic and inelastic partial waves,
 %
  $\,\big|\RE\,a_\ell^\el\big| \leqq \fr{\rh_e^{}}{2}$,~
  $\big| a_\ell^\el\big| \leqq \rh_e^{}$\,,\, and
  $\,\big|a_\ell^\inel \big| \leqq {\sqrt{\rh_i^{}\rh_e^{}}}/2$\,,\,
 %
 where $\,\rh_e^{}$  ($\rh_i^{}$)\, is a symmetry factor of final state
 in $2\to 2$ elastic (inelastic) scattering, and equals
 $1!$ ($2!$) for the final state particles being nonidentical (identical)
 \cite{Dicus:2004rg}.

\begin{figure}[t]
   \begin{center}
     \includegraphics[width=0.6\textwidth]{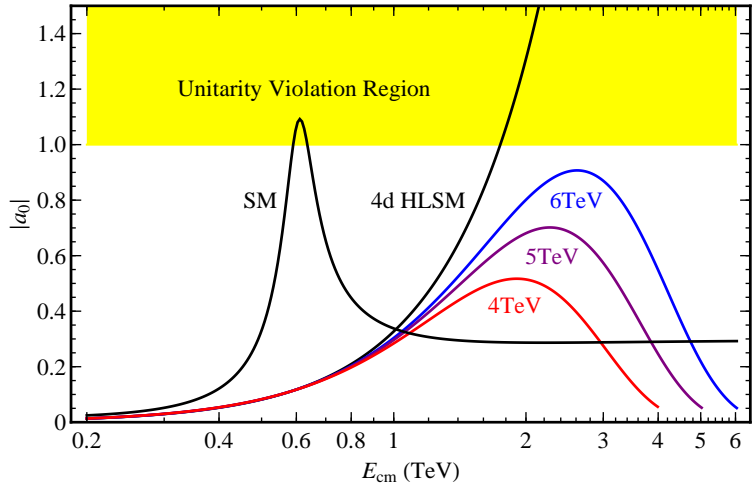}
   \vspace*{-4mm}
   \end{center}
   \caption{Partial wave amplitude of coupled channel scattering
   versus c.m.\ energy $\ECM$.
   Predictions of the HLSM-SDR are shown by (red,\,purple,\,blue) curves
   from bottom to top, for $\,\cut=(4,5,6)$\,TeV.
   For comparison, the amplitudes for the 4d Higgsless SM
   and for the conventional 4d SM (with a 600\,GeV Higgs boson)
   are depicted by the black curves.}
   \label{fig:s-wave}
   \vspace*{-2mm}
\end{figure}

\vspace*{1mm}

 With these, we perform the coupled channel analysis for electrically neutral channels.
 There are two relevant initial/final states, $|W_L^+W_L^-\ra$ and $\fr{1}{\sqrt{2}\,}|Z_L^0Z_L^0\ra$,\,
 and the corresponding amplitudes form a $\,2\times 2$\, matrix,
 \bge
 \label{eq:Tcoup}
   \T_{\text{coup}} ~=~
   \FR{g^2\ECM^2}{8M_W^2}\begin{pmatrix}1+\cos\theta & \sqrt{2}\, \\[1.5mm]
   \sqrt{2} & ~0 \end{pmatrix}.
 \ede
 Thus, we derive the $s$-wave amplitude from the matrix (\ref{eq:Tcoup}) in $n$-dimensions
 and extract the maximal eigenvalue after the diagonalization,
 \bge
 \label{eq:a0-max}
   \big|a_0^{\max}\big| ~=~
   \FR{\tilde g^2}{2^{n+1}\pi^{(n-3)/2}
   \Gamma(\fr{n-1}{2})}\(\FR{\ECM}{2M_W}\)^{\!\!n-2} \,.~~~
 \ede
 Although the partial wave expansion (\ref{eq:PW}) holds for $\,n>3$\,,\,
 we can make analytical continuation of (\ref{eq:a0-max})
 as a function of dimension $\,n\,$ to the full range  $\,2\leqq n\leqq 4$\,.
Here, we perform the analytic continuation on the complex plane of spacetime dimension
$\,n\,$,\, while the one-to-one mapping between $\,n\,$ and $\,\mu\,$ is only defined
within the real interval $\,2\leqq n\leqq 4$\,.

\vspace*{1mm}

 In Fig.\,\ref{fig:s-wave}, we present the unitarity constraint for the standard model
 without a Higgs boson, under Eq.\,(\ref{eq:DFansatz}) with $\gamma =1.5$\,,\,
 where we have varied the transition scale $\cut =(4,\,5,\,6)$\,TeV.
 The shaded yellow region is excluded by the unitarity bound
 $\,|a_0^{\max}| \leqq 1$\,.\,  The $s$-waves
 of HLSM-SDR always have a rather broad ``lump" around $1.5-5$\,TeV
 and then fall off quickly,  exhibiting desired unitary high energy behaviors.
 For comparison, we also show the results of the 4d SM with a {600}\,GeV Higgs boson,
 and the naive 4d Higgsless SM which breaks unitarity at $\,\ECM\simeq 1.74$\,TeV.

\vspace*{1mm}

\begin{figure}[t]
   \begin{center}
     \includegraphics[width=0.48\textwidth,height=6.5cm]{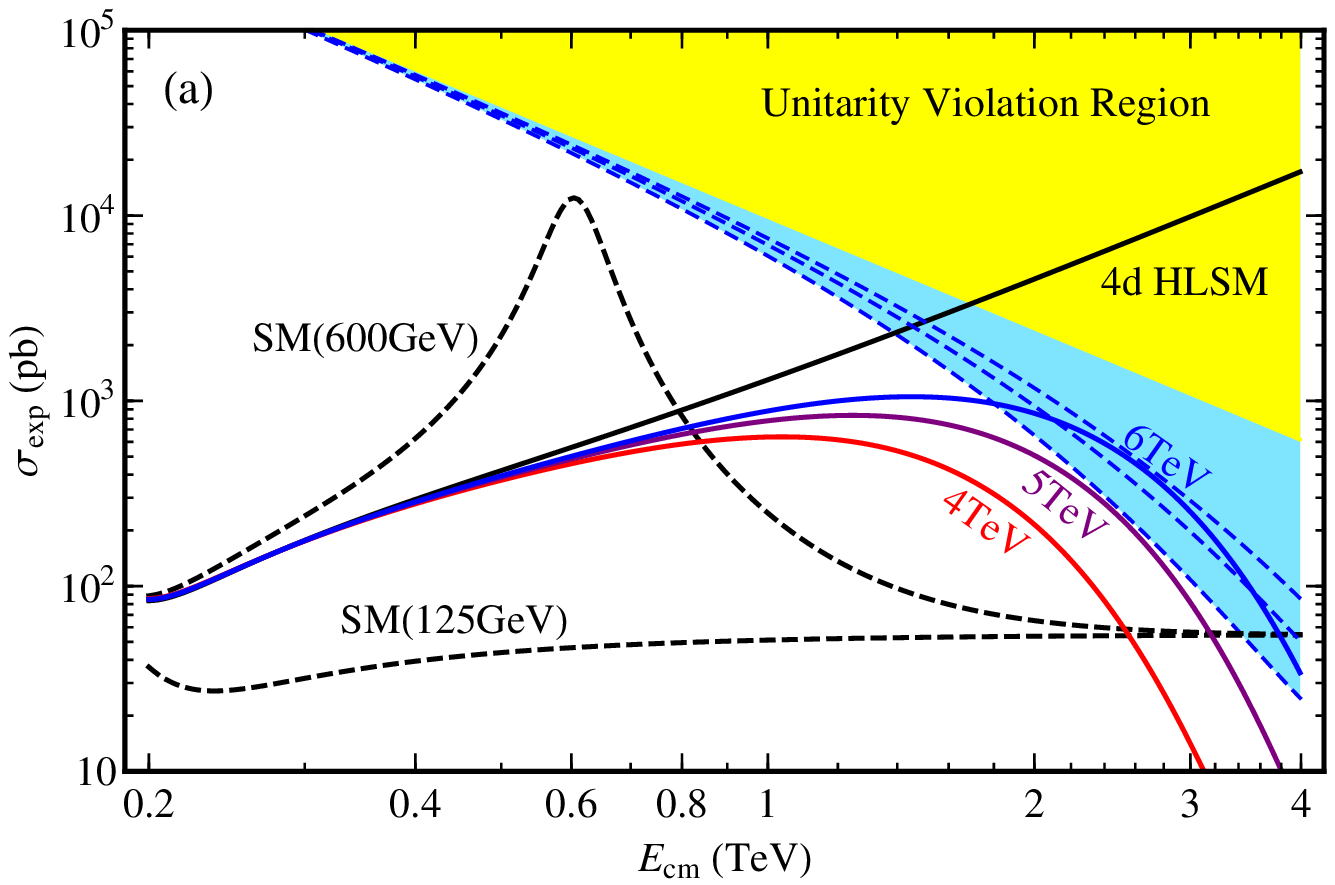}
     \includegraphics[width=0.48\textwidth,height=6.5cm]{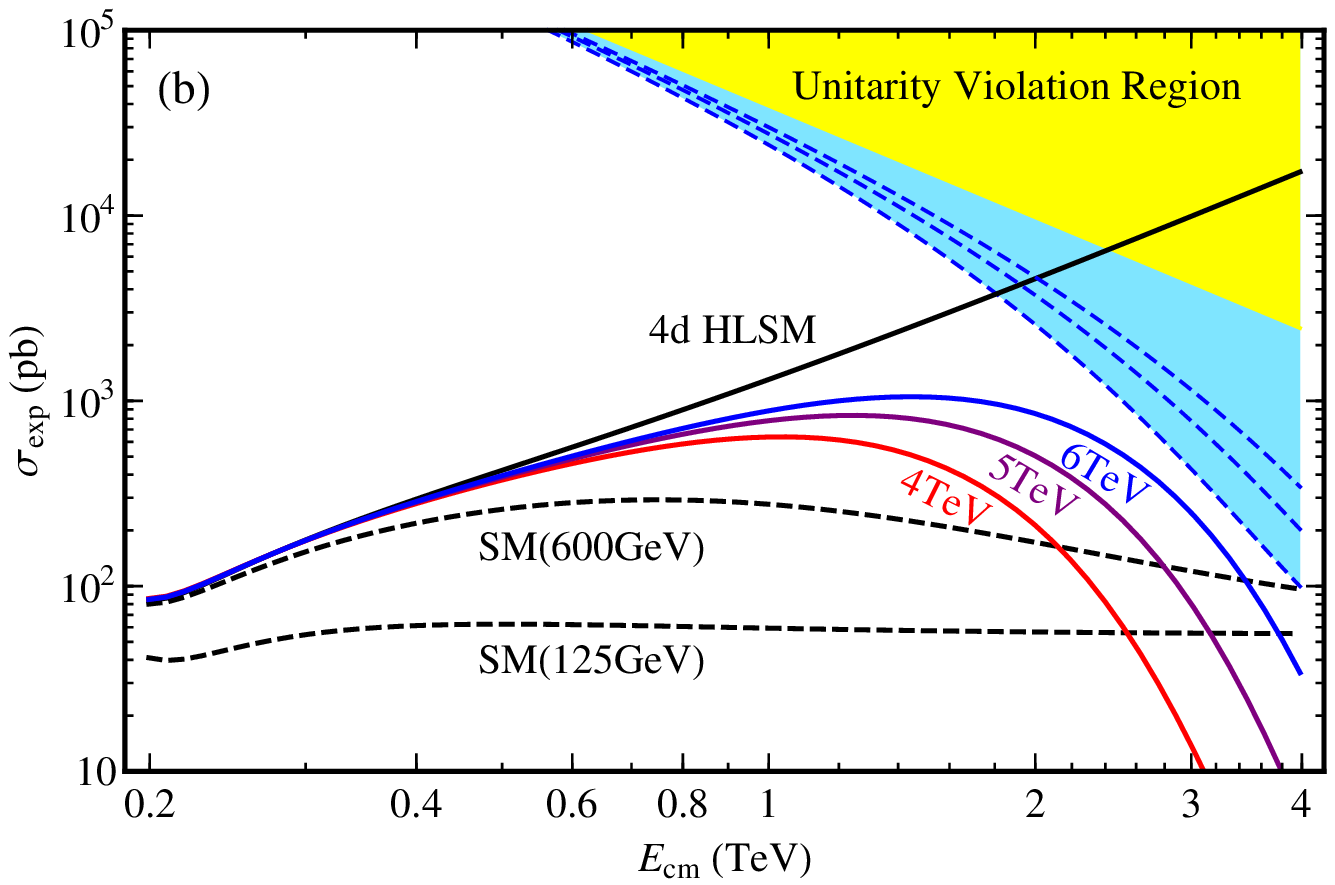}
   \vspace*{-4mm}
   \end{center}
   \caption{Cross sections versus center-of-mass energy $\ECM$ for processes
   (a) $W_L^+W_L^-\to Z_L^0Z_L^0$ and (b) $W_L^+W_L^+\to W_L^+W_L^+$.
   In each plot, predictions of the HLSM-SDR are shown by (red,\,purple,\,blue) curves,
   for $\cut =(4,5,6)$\,TeV. As comparison, results of the conventional 4d SM
   with a light (heavy) Higgs boson of mass $M_h=125$\,GeV (600\,GeV)
   are depicted by black dashed-curves; the result of the 4d Higgsless SM is given
   by black solid-curve. Shaded regions in yellow and light-blue represent
   unitarity violation in 4d and in the HLSM-SDR, respectively. The three blue dashed-lines,
   from bottom to top, show the unitarity bounds for $\,\cut =(4,\,5,\,6)\,$TeV.}
   \label{fig:cs-hl}
 \vspace*{-3.5mm}
 \end{figure}

 Next, we compute the cross sections for
 $W^+_LW^-_L\to Z_L^0Z_L^0$ and $W_L^+W_L^+\to W_L^+W_L^+$, as shown in
 Fig.\,\ref{fig:cs-hl}.  The 4d unitarity condition
 for inelastic cross sections is,
 $\,\si_\inel\leqq 4\pi\rh_e\ECM^{-2}$\,.
 We derive the generalized form in $n$-dimensions,
 \bge
   \label{UniConCS}
   \si_\inel^{} ~\leqq~ \FR{\lam_n^{}\rh_e^{}}{\,4N_0^\nu\ECM^{n-2}\,} \,,
 \ede
 where $\,\lam_n^{}$\, and $\,N_0^\nu$\, are defined below Eq.\,(\ref{eq:PW}).
 Fig.\,\ref{fig:cs-hl} demonstrates how the SDR works as a new mechanism to
 successfully unitarize the high energy behaviors of cross sections
 without invoking extra hypothesized particle (such as the SM Higgs boson).
 Furthermore, the new predictions of the HLSM-SDR are universal and
 show up in all $WW$ scattering channels.
 This is an essential feature of our model and will be crucial for
 discriminating the HLSM-SDR from all other models of the EWSB at the LHC.

\vspace*{1mm}

 Here we note that in $n$-dimensions the cross section $\si$ has its mass-dimensions
 equal $\,[\si]=2-n$\,,\, while the experimentally measured cross section
 $\,\si_{\text{exp}}^{}\,$ always has mass-dimension $-2$\,,\, as the detectors
 record events in 4d. So we need to convert the theory cross section $\si$ under
 the SDR to $\,\si_{\text{exp}}^{}$\,,\,  where the extra mass-dimensions of $\,\si\,$
 should be scaled by the involved energy scale $\,\ECM\,$ of the reaction,
 $\,\si_{\text{exp}}=\si\ECM^{n-4}$\,.

\vspace*{1mm}

 As a final remark, it was found\,\cite{Dicus:2004rg} in 4d
 that varying the phase space may strongly alter the unitarity limit.
 Ref.\,\cite{Dicus:2004rg} observed that the
 {\it enlarged phase space} of $\,2\!\to\! {\cal N}$\, scattering
 (due to properly increasing the number ${\cal N}$ of gauge bosons in the
 final state) will enhance the cross section and result in 
 a new class of much stronger unitarity bounds for all light SM fermions.
 Interestingly, the current study just shows the other way around:
 for $\,2\!\to\! 2$\, scattering,
 {\it reduction of the phase space} of final states
 from decreasing the spacetime dimension $n$ can significantly
 reduce the partial wave amplitudes and cross sections, leading
 to the unitarity restoration.


\vspace*{2mm}
 \section{Weak Boson Scattering in Higgsful SM with SDR}

  \begin{figure}[t]
   \begin{center}
     \includegraphics[width=0.48\textwidth,height=6.5cm]{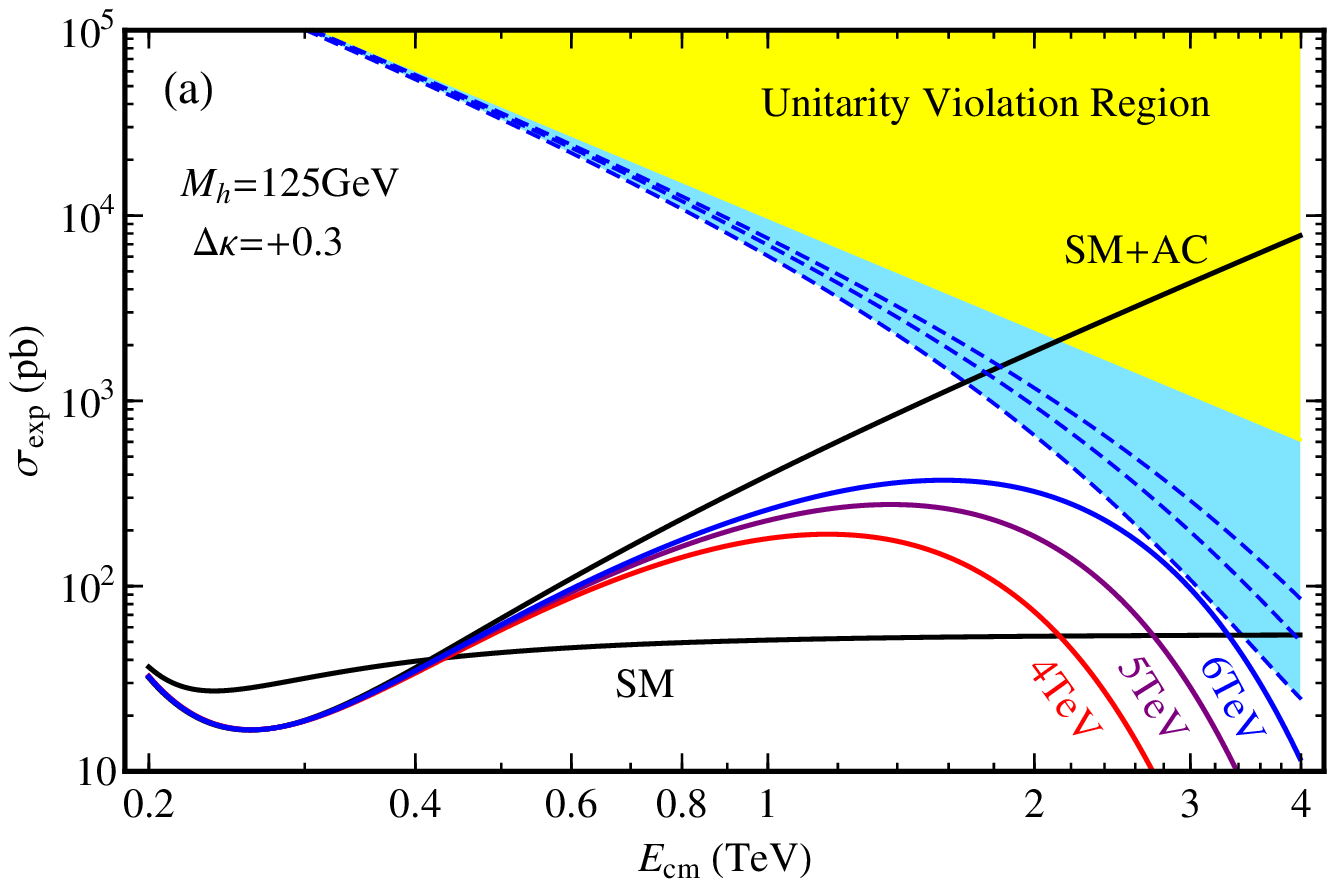}
     \includegraphics[width=0.48\textwidth,height=6.5cm]{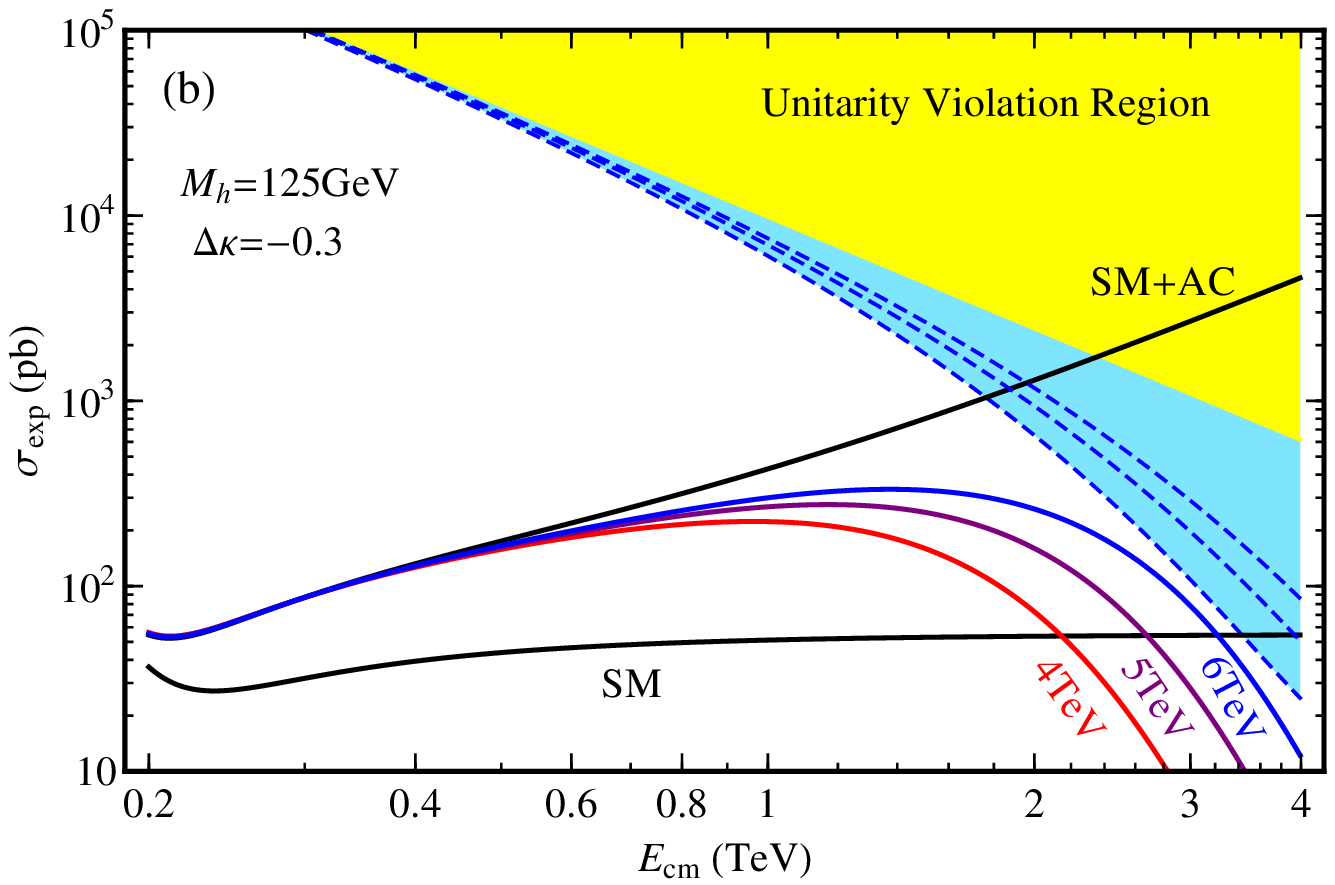}
   \end{center}
   \vspace*{-4mm}
   \caption{Cross sections of $\,W_L^+W_L^-\to Z_L^0Z_L^0$\, versus center-of-mass energy \,$\ECM$\,.\,
   In each plot, predictions of the HFSM-SDR with Higgs mass $M_h=125$\,GeV \cite{LHCnew}
   and anomalous couplings $\Delta\ka=+0.3$ [plot-(a)] and $\Delta\ka=-0.3$ [plot-(b)]
   are shown by (red,\,purple,\,blue) curves, for $\,\cut =(4,5,6)$\,TeV.
   As comparison, results of the conventional 4d SM with $M_h=125$\,GeV
   (labeled by ``SM") and the 4d SM with the same anomalous coupling
   (labeled by ``SM+AC") are depicted by black curves.
   Shaded regions are the same as in Fig.\,\ref{fig:cs-hl}.}
   \label{fig:cs-ac}
 \end{figure}

 In this section, we extend the new mechanism in Sec.\,4
 to the Higgsful SM with SDR (HFSM-SDR).
 In this construction, the quantum-gravity-induced SDR at TeV scales
 provides a natural solution to the hierarchy problem
 that plagues the Higgs boson in the conventional 4d SM.
 For TeV-scale SDR, it is expected that new physics effects induced by the quantum gravity
 will show up in the low energy effective theory. So, the Higgs boson can have
 anomalous couplings with $WW$ and $ZZ$ gauge bosons, and thus behaves as non-SM-like.

\vspace*{1mm}

 In unitary gauge, we can write down the leading anomalous gauge interactions
 of the Higgs boson\,\cite{HVV-0,He:2002qi}
 from the effective Lagrangian (\ref{eq:L-h-Sigma}),
 \beqa
 \dis
 \label{eq:AC}
   \(\Delta\ka vh + \fr{1}{2}\Delta\ka'h^2\)
   \!\left[ \frac{2M_W^2}{v^2}W^+_\mu W^{-\mu}+\frac{M_Z^2}{v^2}Z_\mu Z^\mu \right],~~~
 \eeqa
 where the anomalous couplings $\,\Delta\ka,\,\Delta\ka'\neq 0\,$ represent new physics.
 Besides the hierarchy problem, the conventional 4d SM
 also suffers constraints from the Higgs vacuum instability
 and the triviality of Higgs self-coupling.
 If such a 4d SM would be valid up to Planck scale, then the SM Higgs
 boson mass is bounded within the range\,\cite{Quigg},
 $\,133\,\text{GeV} \lesssim M_h^{} \lesssim 180$\,GeV.\,
 Hence, a Higgs mass outside this window will indicate a non-standard Higgs boson
 in association with {\it new physics.} The Higgs boson in our present model under
 the TeV-scale SDR has anomalous couplings induced from quantum gravity and thus
 behaves as non-SM-like.
 With the recent LHC data\,\cite{LHCnew}, model-independent fits
 already put some interesting constraints on the Higgs anomalous couplings.
 Using the fitting result of \cite{Hfit}, we find that for $\,M_h^{}=125\,$GeV,
 the $\Delta\kappa$ in (\ref{eq:AC})
 is bounded within the range, $\,\Delta\kappa =0.2^{+0.4}_{-0.5}\,$.

\vspace*{1mm}

 In the conventional 4d SM with (\ref{eq:AC}),
 it was found\,\cite{He:2002qi} that the $WW$ scattering has non-canceled large $E^2$ behavior
 and will eventually violate unitarity at TeV scale.
 But in our new model, the TeV-scale SDR can always unitarize $WW$ scattering
 and predicts different behaviors for cross sections, as shown in Fig.\,\ref{fig:cs-ac}
 for $\,\gamma = 1.5\,$.\,
 In Fig.\,\ref{fig:cs-ac}(a)-(b), 
 we study $WW$ scattering process for probing a non-SM Higgs boson
 with mass $\,M_h=125\,$GeV \cite{LHCnew} and 
 sample anomalous couplings $\,\Delta\ka =\pm 0.3\,$.\,
 We find that the cross sections under SDR unitarization (middle colored curves) 
 have sizable excesses above the 4d SM with a 125\,GeV Higgs boson 
 ($\Delta\kappa =0$, flat black curve).
 Then, they fall off in the $2-4$\,TeV region,
 consistent with the corresponding unitarity limits.
 Fig.\,\ref{fig:cs-ac} also shows that for the usual non-unitarized 4d SM
 with nonzero anomalous coupling $\,\Delta\kappa \neq 0\,$,\,
 the cross section (upper black curve)
 rapidly increases and eventually violates unitarity around
 $\ECM = 2$\,TeV for this scattering channel.

\vspace*{1mm}

Before concluding this section, we would like to clarify
the validity range of our effective theory of the SDR.
This validity range lies between the $WW$\,($ZZ$) threshold
(around $160-180$\,GeV) and the UV-cutoff $\,\cut =\order{5\text{TeV}}$\,.\,
It is clearly shown in our Fig.\,\ref{fig:cs-hl}
and Fig.\,\ref{fig:cs-ac},
where the relevant scattering energy $E_{\text{cm}}^{}$
(for our model to be discriminated from the 4d-SM and 4d-HLSM at the LHC)
is always within $0.2-3$\,TeV, which is {\it significantly below
4\,TeV$\,\leqq \cut$\,.}\,
Moreover, within this energy region $0.2-3$\,TeV (relevant to the LHC test),
we can explicitly derive the dimensional flow from Eq.\,(\ref{eq:DFansatz}),
$\,n\simeq 3.98-3.07\,$,\,
(under the typical input of $\,\Lambda_{UV}=5\,$TeV and $\gamma =1.5$), which is
{\it significantly above $\,n=2\,$.}\footnote{Note that only at $n=2$ and its vicinity,
we have $\,E_{\text{cm}}^{}\to\cut\,$ and thus our effective theory should be replaced
by a full theory of quantum gravity.}\,
This clearly shows that for our effective theory study we do not need to invoke
any detailed UV dynamics at or above $\cut$.


\vspace*{2mm}
 \section{Conclusions}

 We have studied the exciting possibility for the onset of 
 spontaneous dimensional reduction (SDR) at TeV scales.
 We demonstrated that the TeV-scale SDR can play a key role to unitarize 
 longitudinal weak boson scattering. We have constructed an effective theory
 of the SM under the SDR, either without a Higgs boson or with 
 a light non-standard Higgs boson. 

\vspace*{1mm}
 
 In the first construction, it nonlinearly realizes the electroweak gauge symmetry 
 and its spontaneous breaking. The model becomes manifestly renormalizable
 at high energies by power counting. We found that the non-canceled $E^2$ contributions 
 to the $WW$ scattering amplitudes are unitarized by the SDR 
 at TeV scales (Fig.\,\ref{fig:s-wave}), and the scattering cross sections
 exhibit different behaviors (Fig.\,\ref{fig:cs-hl}). This will be probed at the LHC.
 Here the recent observation of a 125\,GeV boson at the LHC\,(8\,TeV)
 could be something else, such as a dilaton-like particle\,\cite{dilaton}.
 In passing, we note that the unitarity of $WW$ scattering
 in generic 4d technicolor theories was recently studied in
 Ref.\,\cite{Sannino}.

\vspace*{1mm}

 For the second construction of the Higgsful SM with SDR, we studied the
 $WW$ scattering with a light non-standard Higgs boson of mass $125$\,GeV.
 It has effective anomalous couplings with gauge bosons as induced from
 the TeV-scale quantum gravity effects [cf.\ Eq.\,(\ref{eq:AC})]. 
 Fig.\,\ref{fig:cs-ac}(a)-(b) showed that under the SDR, 
 the cross section of \,$W^+_LW^-_L \to Z_L^0Z_L^0$\, process  
 with anomalous Higgs couplings has distinctive invariant-mass distributions
 from the naive 4d SM Higgs boson over the energy regions around $0.2-3$\,TeV.
 This will be definitively probed by the next LHC runs at 14\,TeV
 collision energy with higher luminosity.

\vspace*{1mm}

 For future works, it is useful to further develop
 a method for quantizing field theories with SDR and compute
 the sub-leading effect of loop corrections in fractional spacetime\,\cite{FST},
 which should have better UV behavior than the usual 4d SM and
 thus is expected to agree even better with the precision data.
 This is fully beyond the current scope and will be
 further explored in future works.
 A systematical expansion of our study in the present Letter is given
 elsewhere \cite{He:2011jv}.

\vspace*{1mm}

As the final remark,
 our effective theory construction is also partly motivated
 by the asymptotic safety (AS) scenario of quantum general relativity (QGR)
 \`{a} la Weinberg \cite{AS-Wein}\cite{AS-rev}.
 In the AS scenario, the theory is originally defined in (3+1)d,
 while solving the exact renormalization group equation of QGR
 points to nontrivial UV fixed point, under which
 the graviton two-point function exhibits
 effective two-dimensional UV behavior\,\cite{AS-rev}.
 Here, the SDR is reflected in anomalous scalings of the fields,
 as well as physical variables like the spacetime curvature.
 Such anomalous scalings share the similarity with our effective theory construction,
 while the field contents are still defined in (3+1)d
 and respect the (3+1)d Lorentz symmetry.
 Our effective theory is a simplified formulation at low energy, so it does not
 rely on any detailed UV dynamics of the AS scenario.
 It is interesting to further study
 the quantitative connection between the SDR and the AS scenario.
 We also note that the Ho\v{r}ava-Lifshitz model \cite{horava} of quantum gravity
 can provide a concrete field-theoretical realization of SDR with UV-completion,
 which has relatively tractable Lagrangian.
 Thus, the various scaling properties in our effective theory are expected
 to arise from the formulation of the Ho\v{r}ava-Lifshitz model.
 We will consider these two interesting scenarios for future works.

\vspace*{6mm}
\noindent
{\bf Acknowledgments}
\\[1.5mm]
We are grateful to Gianluca Calcagni, Steven Carlip and Dejan Stojkovic
for discussing the spontaneous dimensional reduction, to Daniel Litim
for discussing the asymptotic safety, and to Petr Ho\v{r}ava for discussing
the Ho\v{r}ava-Lifshitz gravity.
We thank Francesco Sannino and Chris Quigg
for discussions during their visits to Tsinghua HEP Center.
This work was supported by National NSF of China (under grants 11275101, 11135003,
10625522, 10635030) and National Basic Research Program (under grant 2010CB833000).


\vspace*{3mm}



\begin{thebibliography}{99}

\bibitem{SDR}
For reviews, Steven Carlip,
``Spontaneous dimensional reduction in short-distance quantum gravity",
arXiv:0909.3329 [gr-qc];
``The small scale structure of spacetime", 
arXiv:1009.1136 [gr-qc]; 
and references therein.


\bibitem{TeV-app}
Applications of TeV scale vanishing dimensions to certain astrophysics
and collider phenomenology were recently studied in a different context,
Jonas R.\ Mureika and Dejan Stojkovic,
Phys.\ Rev.\ Lett.\ {106} (2011) 101101 [arXiv:1102.3434 [gr-qc]];
Luis Anchordoqui, De Chang Dai, Malcolm Fairbairn, Greg Landsberg, Dejan Stojkovic,
Mod.\ Phys.\ Lett. A\,{27} (2012) 1250021 [arXiv:1003.5914 [hep-ph]];
Luis A.\ Anchordoqui, De Chang Dai, Haim Goldberg, Greg Landsberg,
Gabe Shaughnessy, Dejan Stojkovic, Thomas J.\ Weiler,
Phys.\ Rev.\ D {83} (2011) 114046 [arXiv:1012.1870 [hep-ph]].


\bibitem{SDR-cosmicray}
T.\ Antoni {\it et al.,} [KASCADE Collaboration],
Phys.\ Rev.\ D\,71 (2005) 072002 [arXiv:hep-ph/0503218];
R.\ A.\ Mukhamedshin, JHEP 05 (2005) 049;
L.\ T.\ Baradzei {\it et al.,} [Pamir Collaboration],
Izv.\ Akad.\ Nauk SSSR, Ser.\ Fiz.\ 50 (1986) 2125
[Bull.\ Acad.\ Sci.\ USSR, Phys.\ Ser.\ 50N11 (1986) 46].


\bibitem{HM}
F.\ Englert and R.\ Brout, Phys.\ Rev.\ Lett.\ {\bf 13} (1964) 321;
P.\ W.\ Higgs, Phys.\ Lett.\ {\bf 12} (1964) 132;
Phys.\ Rev.\ Lett.\ {\bf 13}, 508 (1964); Phys.\ Rev.\ {\bf 145} (1966) 1156;
G.\ S.\ Guralnik, C.\ R.\ Hagen, and T.\ W.\ Kibble,
Phys.\ Rev.\ Lett.\ {\bf 13} (1964) 585.


\bibitem{LHCnew}
G.~Aad {\it et al.,}  [ATLAS Collaboration],
Phys.\ Lett.\ B\,{716} (2012) 1 [arXiv:1207.7214 [hep-ex]].
S.~Chatrchyan {\it et al.,}  [CMS Collaboration],
Phys.\ Lett.\ B\,{716} (2012) 30 [arXiv:1207.7235 [hep-ex]].



\bibitem{dilaton}
For recent studies,
S.\ Matsuzaki and K. Yamawaki,
Phys.\ Rev.\ D\,{86} (2012) 035025 [arXiv:1206.6703];
Phys.\ Rev.\ D\,86 (2012) 115004 [arXiv:1209.2017]; arXiv:1207.5911;
B.\ Bellazzini, C.\ Csaki, J.\ Hubisz, J.\ Serra, J.\ Terning, arXiv:1209.3299;
and references therein.


\bibitem{App}
T.\ Appelquist and C.\ Bernard, Phys.\ Rev.\ D\,{22} (1980) 200.


\bibitem{TreeU}
J.\ M.\ Cornwall, D.\ N.\ Levin,  G.\ Tiktopoulos,
Phys.\ Rev.\ Lett.\ {30} (1973) 1268;
B.\ W.\ Lee, C.\ Quigg,  and H.\ B.\ Thacker,
Phys.\ Rev.\ D {16} (1977) 1519;
Phys.\ Rev.\ Lett.\ {38} (1977) 883.


\bibitem{WW-rev}
For reviews,  Michael S.\ Chanowitz, arXiv:hep-ph/9812215;
Czech.\ J.\ Phys.\ {55} (2005) B45 [arXiv:hep-ph/0412203];
and references therein.


\bibitem{EFT}
For a recent review of the concept of effective field theory,
Steven Weinberg, ``Effective Field Theory, Past and Future",
PoS\,CD\,09 (2009) 001 [arXiv:0908.1964 [hep-th]].


\bibitem{ambjorn}
J.\ Ambj{\o}rn, J.\ Jurkiewicz and R.\ Loll,
Phys.\ Rev.\ Lett.\ {95} (2005) 171301 [arXiv:hep-th/0505113];
Phys.\ Rev.\ D {72} (2005) 064014 [arXiv:hep-th/0505154].


\bibitem{FST}
Gianluca Calcagni,
Phys.\ Rev.\ Lett.\ {104} (2010) 251301 [arXiv:0912.3142];
JHEP {1003} (2010) 120 [arXiv:1001.0571];
JHEP {1201} (2012) 065 [arXiv:1107.5041];
AIP Conf.\ Proc.\ 1483 (2012) 31 [arXiv:1209.1110];
and references therein.


\bibitem{HVV-0}
R.\ Sekhar Chivukula and  V.\ Koulovassilopoulos,
Phys.\ Lett.\ B {309} (1993) 371 [arXiv:hep-ph/9304293].


\bibitem{He:2002qi}
H. J. He, Y.\,P.\ Kuang, C.\,P.\ Yuan and B.\ Zhang,
Phys.\ Lett.\ B {554} (2003) 64 [arXiv:hep-ph/0211229];
and arXiv:hep-ph/0401209, in the proceedings of the
``Workshop on Physics at TeV Colliders", May\,26 - June\,6, 2003,
Les Houches, France.


\bibitem{DRED}
W.\ Siegel, Phys.\ Lett.\ B\,{84} (1979) 193.


\bibitem{ETrev}
For a comprehenive review on this subject,
H. J. He, Y.\,P.\ Kuang and C.\,P.\ Yuan,
DESY-97-056 [arXiv:hep-ph/9704276],
and references therein.


\bibitem{schwinger}
J. Schwinger, Phys.\ Rev.\ {128} (1962) 2425.


\bibitem{deser}
S. Deser, R. Jackiw, and S. Templeton, Ann.\ Phys.\ {140} (1982) 372.


\bibitem{Dicus:2004rg}
Duane A.\ Dicus and Hong-Jian He,
Phys.\ Rev.\ D {71} (2005) 093009 [arXiv:hep-ph/0409131];
Phys.\ Rev.\ Lett.\ {94} (2005) 221802 [arXiv:hep-ph/0502178].


\bibitem{Quigg}
For a review, C.\ Quigg, arXiv:0905.3187 [hep-ph]; and references therein.


\bibitem{Hfit}
E.g., D.\ Carmi, A.\ Falkowski, E.\ Kuflik, T.\ Volansky, J.\ Zupan,
JHEP {\bf 1210} (2012) 196 [arXiv:1207.1718 [hep-ph]];
and references therein.


\bibitem{Sannino}
Roshan Foadi and Francesco Sannino,
Phys.\ Rev.\ D {78} (2008) 037701 [arXiv:0801.0663 [hep-ph]];
Roshan Foadi, Matti Jarvinen, Francesco Sannino,
Phys.\ Rev.\ D {79} (2009) 035010 [arXiv:0811.3719 [hep-ph]].


\bibitem{He:2011jv}
Hong-Jian He and Zhong-Zhi Xianyu, arXiv:1112.1028 [hep-ph].


\bibitem{AS-Wein}
Steven Weinberg, ``Ultraviolet Divergences in Quantum Theories of Gravitation",
in {\it General Relativity: An Einstein Centenary Survey,}
eds.\ S.\ W.\ Hawking and W.\ Israel, Cambridge University Press (1979), p.\,790.


\bibitem{AS-rev}
For a recent review, Daniel F.\ Litim, ``Renormalisation Group and the Planck Scale",
Phil.\ Trans.\ Roy.\ Soc.\ Lond. A\,{369} (2011) 2759 [arXiv:1102.4624 [hep-th]];
and references therein.


\bibitem{horava}
P.\ Ho\v{r}ava, Phys.\ Rev.\ Lett.\ {102} (2009) 161301;
Phys.\ Rev.\ D\,{79} (2009) 084008.


\end{thebibliography}
\end{document}